\documentclass[floatfix,lengthcheck,showpacs,amssymb,amsmath,amsfonts,twocolumn]{aastex62}

\usepackage{lineno}
\usepackage[utf8]{inputenc}
\usepackage{color}
\usepackage{graphicx}
\usepackage{float}
\usepackage{subfigure}
\usepackage{amsmath}
\usepackage{acronym}
\usepackage{multirow}
\usepackage{tikz}
\usetikzlibrary{arrows,shapes,trees,decorations.pathreplacing}
\usepackage{import}
\usepackage{svn-multi}
\usepackage{lineno}
\usepackage{hyperref}
\usepackage{gensymb}
\hypersetup{
    colorlinks=true,
    linkcolor=blue,
    filecolor=magenta,      
    urlcolor=cyan,
}

\newcommand{\mchirpsrc}{\ensuremath{\mathcal{M}^{\mathrm{src}}}}
\newcommand{\chieff}{\ensuremath{\chi_{\mathrm{eff}}}}
\newcommand{\rankingstat}{\ensuremath{\tilde{\Lambda}_{s}}}

\newcommand{\msun}{\ensuremath{\mathrm{M}_{\odot}}}
\newcommand{\past}{\ensuremath{p_{\rm astro}}}
\newcommand{\pastros}{\ensuremath{p_{\mathrm{astro,s}}}}
\newcommand{\bayes}{\ensuremath{B_{c/s}}}
\newcommand{\pastroc}{\ensuremath{p_{\mathrm{astro,c}}}}
\newcommand{\oddsc}{\ensuremath{\mathcal{O}_c}}

\newcommand{\release}{\texttt{\url{www.github.com/gwastro/single-search}}}

\begin{document}
\title[]{A Search for Gravitational Waves from Binary Mergers with a Single Observatory}

\correspondingauthor{Alexander H. Nitz}
\email{alex.nitz@aei.mpg.de}

\author[0000-0002-1850-4587]{Alexander H. Nitz}
\affil{Max-Planck-Institut f{\"u}r Gravitationsphysik (Albert-Einstein-Institut), D-30167 Hannover, Germany}
\affil{Leibniz Universit{\"a}t Hannover, D-30167 Hannover, Germany}

\author[0000-0003-1354-7809]{Thomas Dent}
\author[0000-0002-4289-3439]{Gareth S. Davies}
\affil{Instituto Galego de F\'{i}sica de Altas Enerx\'{i}as, Universidade de Santiago de Compostela, Santiago de Compostela, Galicia, Spain }

\author[0000-0002-5304-9372]{Ian Harry}
\affil{University of Portsmouth, Portsmouth, PO1 3FX, United Kingdom}
\affil{Kavli Institute of Theoretical Physics, UC Santa Barbara, CA}


\begin{abstract} 
We present a search for merging compact binary gravitational-wave sources that produce a signal appearing solely or primarily in a single detector. Past analyses have heavily relied on coincidence between multiple detectors to reduce non-astrophysical background.
However, for $\sim40\%$ of the total time of the 2015-2017 LIGO-Virgo observing runs only a single detector was operating. We discuss the difficulties in assigning significance and calculating the probability of astrophysical origin for candidates observed primarily by a single detector, and suggest a straightforward resolution using a noise model designed to provide a conservative assessment given the observed data. 
We also describe a procedure to assess candidates 
observed in a single detector when multiple detectors are observing. 
We apply these methods to search for binary black hole (BBH) and binary neutron star (BNS) mergers in the open LIGO data spanning 2015-2017.  The most promising candidate from our search is 170817+03:02:46UTC (probability of astrophysical origin $p_{\rm astro} \sim 0.4$): if astrophysical, this is consistent with a BBH merger with primary mass  $67_{-15}^{+21}\,\msun$,
suggestive of a hierarchical merger origin.
We also apply our method to the analysis of GW190425 and find $\past{} \sim 0.5$, though this value is highly dependent on assumptions about the noise and signal models.

\end{abstract}

\section{Introduction}

To date the Advanced LIGO~\citep{TheLIGOScientific:2014jea} and Virgo~\citep{TheVirgo:2014hva} observatories
have observed over a dozen binary black hole (BBH) mergers~\citep{LIGOScientific:2018mvr,Nitz:2018imz,Nitz:2019hdf,Venumadhav:2019tad}. 
These BBH mergers were found in time when multiple observatories were operating by algorithms requiring coincident detection of a source by multiple detectors. 
Several dozen additional observations are expected from the recent third observing run\footnote{https://gracedb.ligo.org/superevents/public/O3}.

Coincident observation of gravitational-wave signals has been a staple method for gravitational-wave detection and is employed by several low-latency~\citep{Nitz:2018rgo,Messick:2016aqy,sachdev2019gstlal,Adams:2015ulm,Hooper:2011rb} and  
archival analyses~\citep{Messick:2016aqy,sachdev2019gstlal,Usman:2015kfa,Venumadhav:2019tad,Klimenko:2015ypf,Babak:2012zx,Allen:2005fk}. A key advantage of requiring coincident observation is the ability to discard non-astrophysical candidates which cannot be excluded based on signal morphology or environmental monitoring alone. LIGO and Virgo data is known to contain non-astrophysical noise transients~\citep{Nuttall:2015dqa,TheLIGOScientific:2016zmo,TheLIGOScientific:2017lwt,Cabero:2019orq}. As the cause of much of the transient noise is unknown, it cannot easily be
excluded with high confidence that an observation by a single detector is due to an instrumental source. Signals from astrophysical sources are constrained by 
detector locations and antenna patterns; they appear in multiple detectors with predictable times of arrival and signal strengths. If coincident observations are required, the rate of false coincidences can be empirically estimated by time-shifting the data from one detector by an amount greater than the inter-site time-of-flight. This method can estimate the false alarm rate to a precision of 1 in $10^3-10^5$ years. In contrast, a single-detector analysis cannot empirically estimate the rate of false alarms at a rate less than 1 per observation time. 

Several archival analyses have included single-detector LIGO data in support of joint multi-messenger detection \citep{Nitz:2019bxt,Magee:2019vmb,Authors:2019fue,Stachie:2020kqz}, and despite the difficulties in rejecting noise transients with high confidence, two methods to identify gravitational-wave candidates in a single detector have already been employed in production 
low-latency analyses \citep{Nitz:2018rgo,Messick:2016aqy}\footnote{The detection method of \citet{Nitz:2018rgo} was used for online LIGO/Virgo analysis during O2, but was not employed in O3}. The GstLAL-based analysis of \citet{Messick:2016aqy} was responsible for the initial identification of GW170817 and GW190425.
Candidates were initially assessed for issuing alerts for astronomical follow-up~\citep{LIGOScientific:2019gag} either by applying a pre-determined threshold~\citep{Nitz:2018rgo} or by background extrapolation~\citep{Messick:2016aqy}. Neither method is suitable for assigning a robust candidate false alarm rate or probability of astrophysical origin \past. 

In this paper, we 
describe a method to detect and assess significance for gravitational-wave candidates identified by only a single detector. We analyze the entire set of public LIGO data from 2015-2017 \citep{Vallisneri:2014vxa,Abbott:2019ebz} using this method and produce a ranked list of BBH and BNS candidates from each detector, including times when multiple detectors were observing.  We focus on the analysis of data from the two LIGO instruments due to the significantly larger observing time and sensitivity. For candidates identified by only a single detector at a time when multiple detectors are observing, we combine our prior single-detector odds with support from the additional detectors, including Virgo when available. This may be useful for identifying candidates in coincident time where their signal is marginal in all but one detector. A similar approach to assessing such candidates was investigated in \cite{Zackay:2019btq}.

Our search does not find any new BBH candidates with $\past > 0.5$, nor do we find any new BNS candidates of interest. During the first and second observing runs of Advanced LIGO and Virgo there were 173 days of time with multiple detectors observing simultaneously, and an additional 115 days where either LIGO-Hanford or LIGO-Livingston were operating alone. Given the rate of mergers, we would expect
2-3 sources in single-detector time identifiable with a false alarm rate of ${\sim}1$ per observation time; this expected count is consistent with our results within
$2\sigma$.

The most significant candidate from our search is 170817+03:02:46UTC which occurred during triple-detector observing time and we estimate has $p_{\rm astro} \sim 0.4$. This candidate was earlier reported in \citet{Zackay:2019btq}. 
If astrophysical, the source may be consistent with a BBH with one or both components formed through hierarchical merger \citep{Gayathri:2019kop}. 
However, establishing the presence of a hierarchical formation channel requires assessing this candidate in the context of the full observed population, which we do not do here; see \citet{Fishbach:2019ckx} for further discussion of high-mass BBH population outliers. 

We focus on statistical statements for single-detector events using methods designed for initial identification of signals in gravitational wave data, i.e.\ search algorithms; however, we expect other \emph{followup} methods to contribute substantially to the final assessment of the possible astrophysical origin of such events.  Such followups are broadly of two types: first, checks on the validity of the data and stability of detector operation around the time of the event (for instance \citealt{TheLIGOScientific:2016zmo}); second, detailed checks of consistency of the strain data with expectations for a gravitational-wave signal described by general relativity plus typical noise realizations.  

Followups for detector state and data quality are routinely carried out by LIGO-Virgo and require auxiliary information not currently in the public domain.  Detailed signal consistency checks may be performed by examining the residuals of the data after removing best fit signal waveform models \citep{Nielsen:2018bhc,LIGOScientific:2019hgc}, which we perform here for our BBH candidate events, as well as by comparing the inferred source parameters to astrophysical expectations, and by explicitly testing for evidence of deviations of the signal from the form expected in general relativity \citep{GW150914:TGR}.

\section{Searching for Binary Mergers}

We first briefly summarize the process of identifying and assessing possible candidates, before giving a more detailed account of each stage. Possible single-detector candidates are identified using matched filtering 
implemented in the PyCBC search \citep{pycbc-github,Usman:2015kfa}. These candidates are ranked in a similar manner to that used for the 2-OGC analysis \citep{Nitz:2019hdf}. Each candidate is then assigned a probability of astrophysical origin based on models of the signal and noise distributions.  Finally, if data from additional detectors are available, we update the \past\ value based on the support that an astrophysical signal is present in the additional detectors. 

\subsection{Candidate Selection}

\begin{figure}[tb!]
    \centering
    \includegraphics[width=\columnwidth]{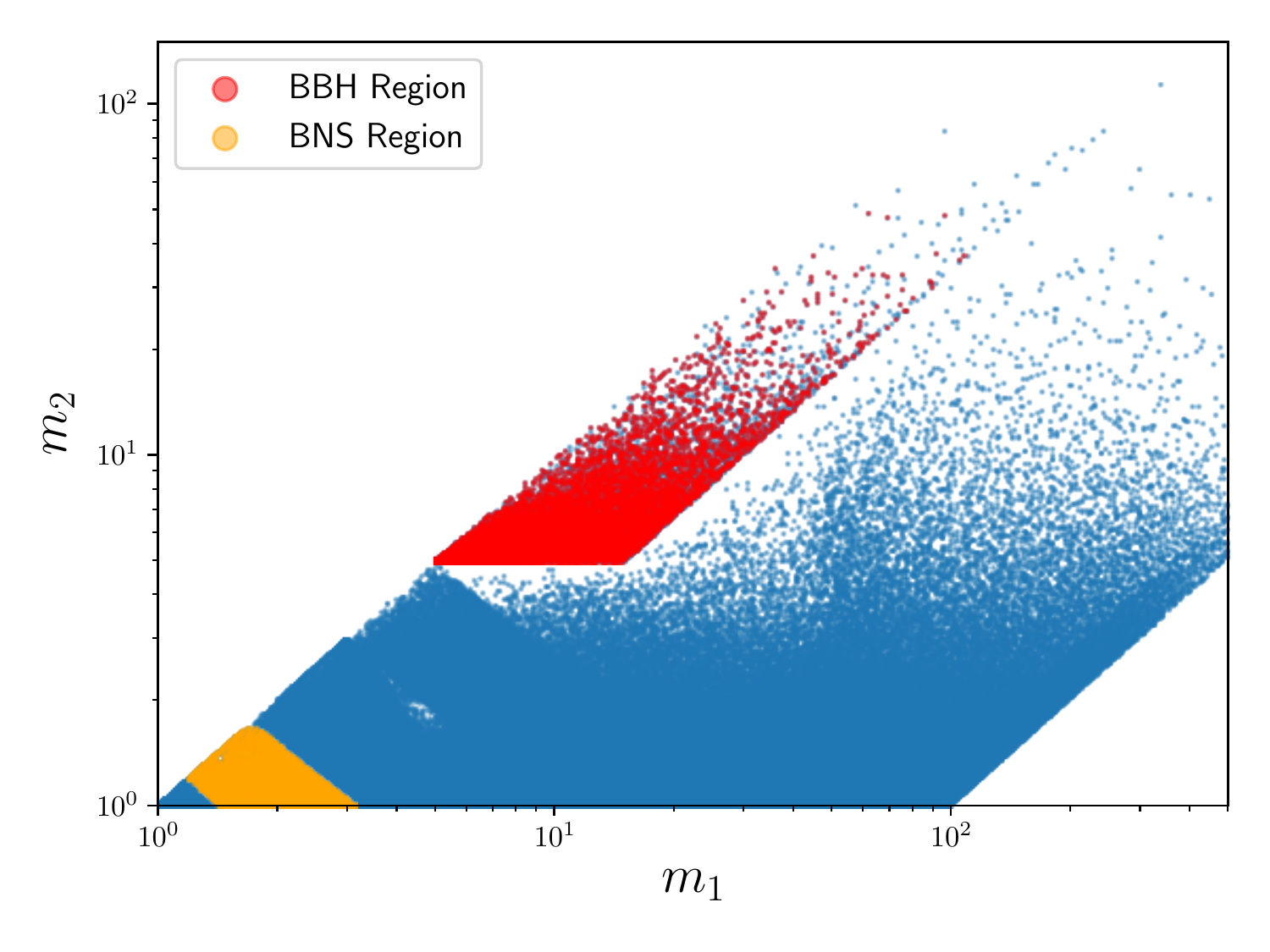}
    \caption{Regions of the 2-OGC template bank selected as BBH-like (red) or BNS-like (orange) where our analysis considers single-detector candidates. All published mergers have come from these two regions. Combined, these two regions contain ${\sim}9\%$ of the full template bank.}
    \label{fig:bank}
\end{figure}

After a target range of source parameters in component masses and spins is chosen, a bank of discrete gravitational waveform templates is placed to ensure sensitivity over this range; for methods used in the analysis presented here, see \citet{DalCanton:2017ala,Brown:2012nn,Ajith:2012mn}. 
Each template is correlated against the gravitational-wave data from each detector, producing a set of signal-to-noise (SNR) time series~\citep{Allen:2005fk}. Peaks in these time series are identified using the procedure of \citep{Nitz:2019hdf} and recorded as single-detector candidates. 

We then select single-detector candidates 
from regions of the template bank used in the 2-OGC analysis \citep{DalCanton:2017ala} corresponding to BBH or BNS candidates, as shown in Fig~\ref{fig:bank}. The BBH region is similar to the one defined in \cite{Nitz:2019hdf}, defined by $m_{1,2} > 5~\,\msun$, $q<3$, $\mathcal{M} < 60\,\msun$, where $m_1$ and $m_2$ are the masses of the primary and secondary components, respectively, $q$ is the mass ratio ($m_1/m_2$), and $\mathcal{M}$ is the chirp mass. An excess rate of candidates is observed at high effective spin (both aligned and anti-aligned), while none of the existing confident detections exhibit high effective spin~\citep{LIGOScientific:2018mvr,Zackay:2019tzo,Nitz:2019hdf,Huang:2020ysn}. To reduce background, and restrict ourselves to the most likely region to make detections, we reduce the range of the effective spin to $|\chieff| < 0.8$.

As the BNS space is significantly less constrained by observation we inform our bounds using \cite{Ozel:2012ax} for BNS
and impose chirp mass and $\chieff$ boundaries, similarly to \citet{Nitz:2019bxt}. 
Based on the recent observation of the massive BNS merger GW190425 ($\mathcal{M} \sim 1.4\,\msun$, \citealt{Abbott:2020uma}) the upper mass boundary was raised from 1.36 (as in ~\cite{Nitz:2019bxt,Nitz:2019hdf}) to $1.5\,\msun$, thus we include systems with $1.06 < \mathcal{M} < 1.5\,\msun$ and $|\chieff| < 0.2$. 

Candidates during time flagged as affected by instrumental artefacts in the data release are removed from the analysis~\citep{TheLIGOScientific:2016zmo,TheLIGOScientific:2017lwt,Vallisneri:2014vxa,Abbott:2019ebz}. However, unflagged transient noise does remain in the data, and can produce large matched-filter SNR values \citep{Nuttall:2015dqa,TheLIGOScientific:2016zmo,TheLIGOScientific:2017lwt,Cabero:2019orq}. To further suppress non-Gaussian transient noise we apply two signal-consistency tests~\citep{Nitz:2018rgo,Allen:2004gu}.
The tests produce $\chi^2$ distributed values if data contains Gaussian noise and a possible signal matching the template; the most likely value of $\chi^2$ per statistical degree of freedom (reduced chi-squared, $\chi^2_{r}$) is unity when the data contains Gaussian noise or Gaussian noise and a signal.  We then discard candidates with $\chi^2_{r} > 1.5$ for our primary time-frequency test \citep{Allen:2004gu}
and $\chi^2_{r,sg} > 4$ for the statistic defined in \citet{Nitz:2018rgo}. The surviving candidates are then ranked as described in the next section.

\subsection{Single-detector Candidate Ranking}

We construct our ranking statistic $\Lambda_{s}$ separately for each detector in an analogous manner to the multi-detector statistic derived in \citet{Nitz:2019hdf,Davies:2020,Simone:2020}. Our ranking statistic is constructed from models of the rate density of noise and signal events at a given ranking statistic value.

The first step is to calculate the re-weighted SNR statistic $\hat{\rho}$ \citep{Babak:2012zx} via
\begin{equation}
 \hat{\rho} = \begin{cases} 
        \rho & \mathrm{for}\ \chi^2_r \leq 1, \\
        \rho\left[ \frac{1}{2} \left(1 + \left(\chi^2_r\right)^3\right)\right]^{-1/6} & 
        \mathrm{for}\ \chi^2_r > 1,
    \end{cases}
\end{equation} 
where $\chi^2_{r}$ is the signal consistency test of \citet{Allen:2004gu} and $\rho$ is the matched-filter SNR. 
We expect the relative densities of signal and noise events over the $\rho,\chi^2_r$ plane to be well described by a function of $\hat{\rho}$. 

As in \citet{Nitz:2019hdf,Simone:2020} we scale this re-weighted SNR by an estimate of the short-term variation in the power spectral density (PSD), $v_S(t)$. This
accounts for frequency-independent changes in the PSD over tens of seconds. Some classes of environmental or other disturbances to the detectors produce broadband noise, which can be down-weighted by this correction \citep{Simone:2020}. A similar approach was also introduced in \citet{Zackay:2019kkv}.
The PSD variation correction is applied as \citep{Nitz:2019hdf}:
\begin{equation}
     \tilde{\rho} = \hat{\rho} v_S(t)^{-0.33}. 
\end{equation}

From our distribution of candidates we estimate the noise rate using a fitted model. 
For each template labelled by $i$ we count the number of candidates with $\tilde{\rho}>6$, and fit the distribution of $\tilde{\rho}$ values for candidates with $6 < \tilde{\rho}< 8$ to an exponential \citep{Nitz:2017svb}. We then smooth the fitted values over the set of templates parameterized by template duration and effective spin using a Gaussian kernel. This smoothing significantly reduces the variance due to small number statistics for individual template fits, while accounting for variation over the parameter space. Our estimate of the noise rate $R_{N,i}$ for each template as a function of $\hat{\rho}$, using these fits, is
\begin{equation}
\label{eqn:3}
R_{N,i}(\tilde{\rho}) = r_{i} e^{-\alpha_{i}(\tilde{\rho}-\rho_{\rm th})}
\end{equation}
where $r_{i}$ and $\alpha_i$ are the amplitude and slope of the exponential fit, respectively, and $\rho_{\rm th}$ is the fit threshold value, here equal to $6$. Note that this procedure treats the noise rate as a constant over time, which may be sub-optimal if the detectors are in considerably different configurations. A future improvement may be to allow time variation in the fit parameters.

We also use a signal rate model which accounts for the sensitivity of the instrument at the time of a candidate, similarly to \cite{Davies:2020}, and may incorporate a model of the mass distribution of detected signals over search templates, as in \cite{Nitz:2019hdf,Dent:2013cva}.

The signal model for BBH, giving the rate density of signals per template at a given statistic $\tilde{\rho}$, is written
\begin{equation}
    \label{eqn:4}
    R_{S, i} = \left(\frac{\sigma_i}{\bar{\sigma}_i}\right)^3 \left(\frac{\mathcal{M}_i}{\mathcal{M}_{\rm ref}}\right)^{11/3},
\end{equation}
where $\sigma_i$ is the (time-varying) noise-weighted amplitude of the $i$'th template \citep{Allen:2005fk} and $\bar{\sigma}_i$ its average over the data analyzed for the given detector. 
(We omit an overall constant from $R_{S,i}$ corresponding to the actual rate of astrophysical signals in the detector, which we do not model here.) 
$\mathcal{M}_i$ is the chirp mass of the template associated with a given candidate, scaled by a reference chirp mass $\mathcal{M}_{\rm ref} = 20\, \msun$.  This chirp mass dependence results from modelling the signal rate density as constant over $\mathcal{M}$ and observing that the density of templates (and hence of noise events) in the BBH region varies as ${\sim}\mathcal{M}^{-11/3}$.  
For BNS candidates we omit the second term weighting by chirp mass, which instead implies a constant prior probability of signal per template. A more sophisticated BNS mass model can be employed in our method when the population is better understood.

Single-detector candidates are then ranked for each detector by the ratio of signal and noise rate densities as
\begin{equation}
    \Lambda_s = \log(R_S) - \log(R_N),
\end{equation}
where each term is a function of the candidate's ranking statistic $\tilde{\rho}$, its template $i$, and of the detector via the noise model fits and the time-dependent sensitivity $\sigma_i$, as detailed in equations \ref{eqn:3} and~\ref{eqn:4}. A merger signal or noise transient may produce multiple, correlated candidates at the same time: thus within a sliding 10\,s window, we keep only the one with the largest value of $\Lambda_s$. We can thus approximate the appearance of signal and noise candidates on timescales ${\gg}10$\,s as Poisson processes.

\subsection{Probability of Astrophysical Origin}

To estimate the probability of astrophysical vs.\ terrestrial origin of a single-detector candidate, we need to know the actual rate of astrophysical signals compared to noise events at the candidate's ranking statistic $\Lambda_s$. While 
$\Lambda_s$ aims to approximate the ratio of signal to noise rates (up to the addition of a constant dependent on the true signal rate and on average detector sensitivity), we do not assume that the functional forms used are complete or accurate models of the observed signal and noise processes. 
Instead, as is done for standard coincident searches, we determine the signal distribution over our ranking statistic by analysing a large number of simulated signals added to real data. 

Our noise rate distribution, however, cannot be empirically estimated for values of $\Lambda_s$ which have not been sampled. At small $\Lambda_s$ we expect the great majority of candidates will be of noise origin, thus the actual distribution will be a good approximation of the noise; however at large values we will have very few noise samples per analysis time, moreover any actual candidates may have a nonzero probability of signal origin. 
Instead we must use a model of the noise rate at high $\Lambda_s$; we aim to choose a model that is not overly optimistic when considering the observed data. 

Then using models of the noise and signal rates which are derived in Sections~\ref{sec:sm} and~\ref{sec:nm}, respectively, the probability that a given candidate is astrophysical is 
\begin{equation}
    \pastros = \frac{\mu_S(\Lambda_s)}{\mu_S(\Lambda_s) + \mu_N(\Lambda_s)}
\end{equation}
where $\mu_S$ and $\mu_N$ are the rate of signal and noise single-detector candidates at a given ranking statistic, respectively, expressed as densities over $\Lambda_s$ \citep{Farr:2015,Abbott:2016nhf}.

\subsubsection{Signal model}
\label{sec:sm}
Our simulated signal population assumes a distribution that is isotropically distributed in the sky and binary orientation and uniform in volume. To determine the distribution of signals over $\Lambda_s$, we analyze $O(10^4)$ simulated sources. The distribution of masses and spins are chosen from within the searched region. For our BBH analysis, the signals are taken to be uniform in chirp mass, mass ratio ($q \in [1, 3]$), and spin magnitude. We expect the foreground statistic distribution to be largely independent of the exact details of this distribution~\citep{Schutz:2011tw}. The signals are added to real data and we smooth the resulting distribution of associated ranking statistic values using a KDE with bandwidth 1. The overall amplitude of the signal rate $\int \mu_S\, d \Lambda_s$ is fixed in each detector using the confident mergers identified in previous coincident analyses \citep{LIGOScientific:2018mvr,Nitz:2019hdf,Venumadhav:2019tad}. 
We count the number of known gravitational-wave sources in a given detector with single-detector SNR greater than 8. This amounts to 8 (5) confident coincident signals that were observed above ranking statistic threshold of 6.5 (9) for LIGO-Livingston (LIGO-Hanford). The expected rate of signals in each detector is then fixed to the rate of previously observed signals above the ranking statistic threshold. This determination of the signal rate is subject to Poisson (counting) uncertainty, however this uncertainty has less impact on our assessment of the most significant candidates than the uncertainty in the noise distribution from the choice of noise model. 

\subsubsection{Noise model}
\label{sec:nm}

The largest hurdle in assessing the astrophysical probability of a single-detector candidate arises from the inability to empirically estimate the noise density for the most significant candidates. 
\begin{figure}[tb!]
    \centering
    \includegraphics[width=\columnwidth]{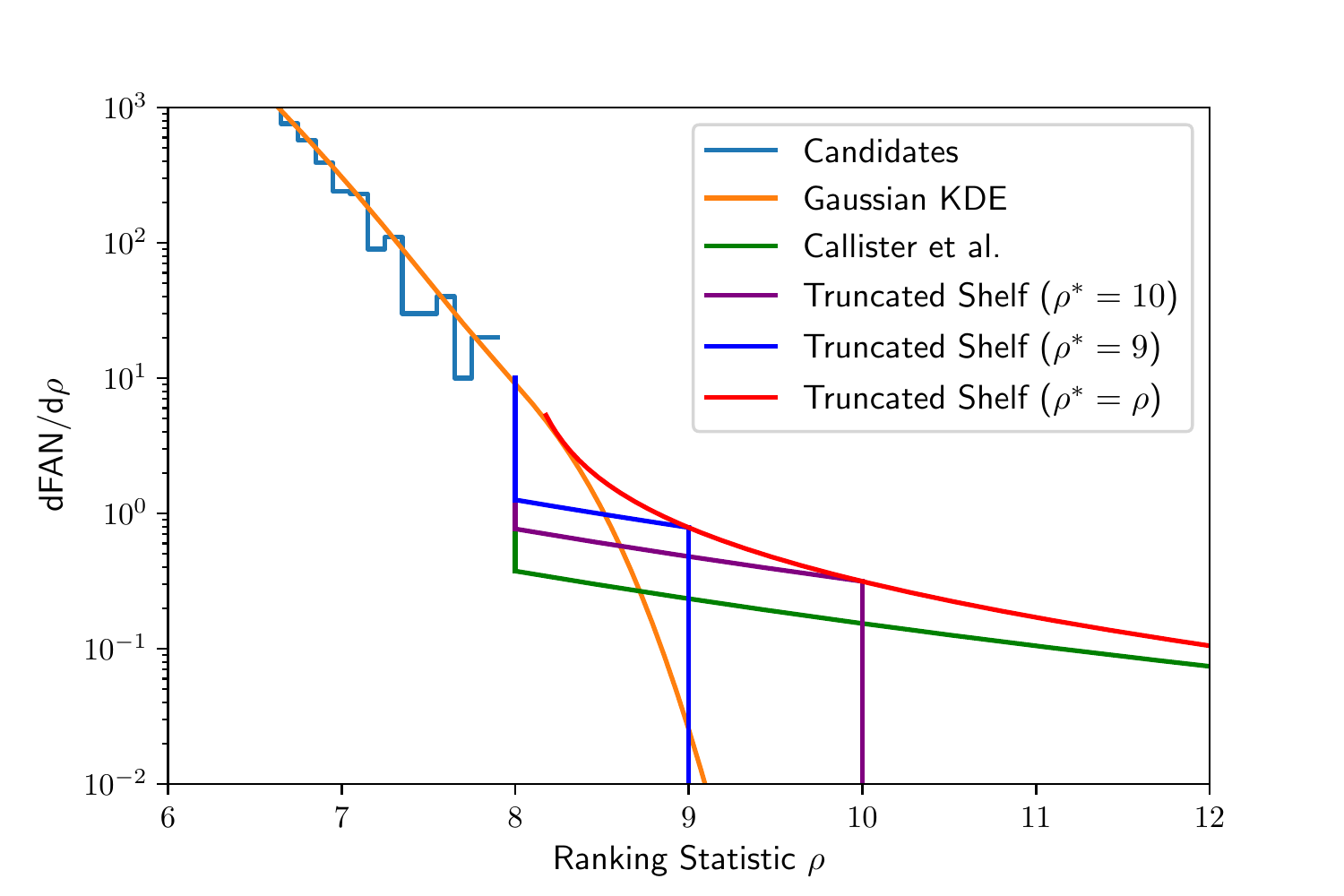}
    \caption{Comparison of noise models using samples drawn from an idealized analytic distribution ($\chi$ distribution with 2 degrees of freedom). The distribution of ranking statistic values $\rho$ for the observed candidates 
    are shown in blue, along with a Gaussian KDE (orange) with  bandwidth 1.
    The noise model proposed in \citet{Callister:2017urp}, normalised such that a single noise count is expected with $\rho > 8$ is shown in green. This can be compared to our suggested `truncated shelf' noise model, where the distribution is shown for an example candidate at $\rho_{\rm max}=9$ (blue) and $\rho_{\rm max}=10$ (purple). The red curve shows the noise density of our model for an arbitrary ranking statistic value. For large values of $\rho$, our model approaches the \citet{Callister:2017urp} model. The vertical axis is normalised so that when integrated over $\rho$ the result is candidates over the total observation time $T$, which for noise represents a false alarm number $\mathrm{FAN} \equiv T\cdot \mathrm{FAR}$. 
    }
    \label{fig:model}
\end{figure}
At a sufficiently low SNR, the bulk of the distribution is from noise and the density of candidates is high enough to be readily estimated via standard methods. 
However, to determine the noise density for the tail of the distribution, where we expect confident observations of single-detector events, a method of extrapolation is required to extend the noise model.

A KDE approach to extrapolate the noise density has been 
employed by the GstLAL-based compact binary searches \citep{Messick:2016aqy}. While this approach has been successful in identifying single-detector events from BNS sources \citep{TheLIGOScientific:2017qsa, Abbott:2020uma}, the predicted noise density is highly dependent on the parameterization and kernel chosen. 
Although one can expect the population at low SNR to be dominated by contributions from Gaussian noise, there is no evidence to support that noise would continue any particular distribution for larger SNR candidates. Non-Gaussian noise transients are known to occur \citep{Nuttall:2018xhi,Cabero:2019orq} and novel transients may only occur on the order of once per observation time. This leads to the possibility of over-estimating the significance of candidates in the tail of the noise distribution by an unknown factor. 

A second approach was proposed in \citet{Callister:2017urp} which addresses the issue of overestimating significance at high SNR by suggesting a noise model which is the ``flattest'' possible (i.e.\ the noise distribution decreases least with increasing SNR) while retaining the property that louder signals are no less significant than quieter ones. The resulting noise model has the same functional form as the expected signal distribution, but the expectation for the number of noise candidates with ranking statistic greater than the second-loudest candidate
is set to a constant: 3 in \citet{Callister:2017urp}, 1 in this work. A drawback of this model is evident in Fig.~\ref{fig:model}, where this model is compared to a KDE for samples drawn from a simple analytic distribution ($\chi$ distribution with 2 degrees of freedom). 
While this model may be conservative in the limit of large ranking statistic, we see that it has a sharp discontinuous drop just above the well-measured bulk distribution. The Callister method would underestimate the noise by up to 1.5 orders of magnitude with respect to a KDE for candidates with $\rho$ between 8 and 8.7.

We employ a third approach, which
is conservative with respect to these two models while generally retaining the characteristic that louder sources not be less significant than quieter ones. Similarly to \citet{Callister:2017urp} we assume the noise probability density function is proportional to the signal distribution at high SNR; however, to normalise the distribution we truncate it \emph{at the ranking statistic of the candidate event}, instead of extending to infinity: in each case, we normalise these distributions so that $1$ noise event is expected between the ranking statistic of any given candidate and the next candidate in decreasing statistic order.

This `truncated shelf' method avoids the precipitous drop in noise density observed in the Callister model (green) of Fig.~\ref{fig:model}, and simulates the presence of glitches with SNR (or ranking statistic) values bounded above. 
The resulting noise density estimates for candidates with ranking statistic values $\rho^*$ of 9 and 10 are shown by the blue and purple curves, respectively. 
The red curve then shows the noise density assigned by our model to candidates with arbitrary ranking statistic value.

For large values our model approaches that of \citet{Callister:2017urp}. Thus, the `truncated shelf' density estimate, while suitable for evaluating the few loudest most candidates, may be overly conservative when considering a larger population. 

\begin{figure}
    \centering
    \includegraphics[width=\columnwidth]{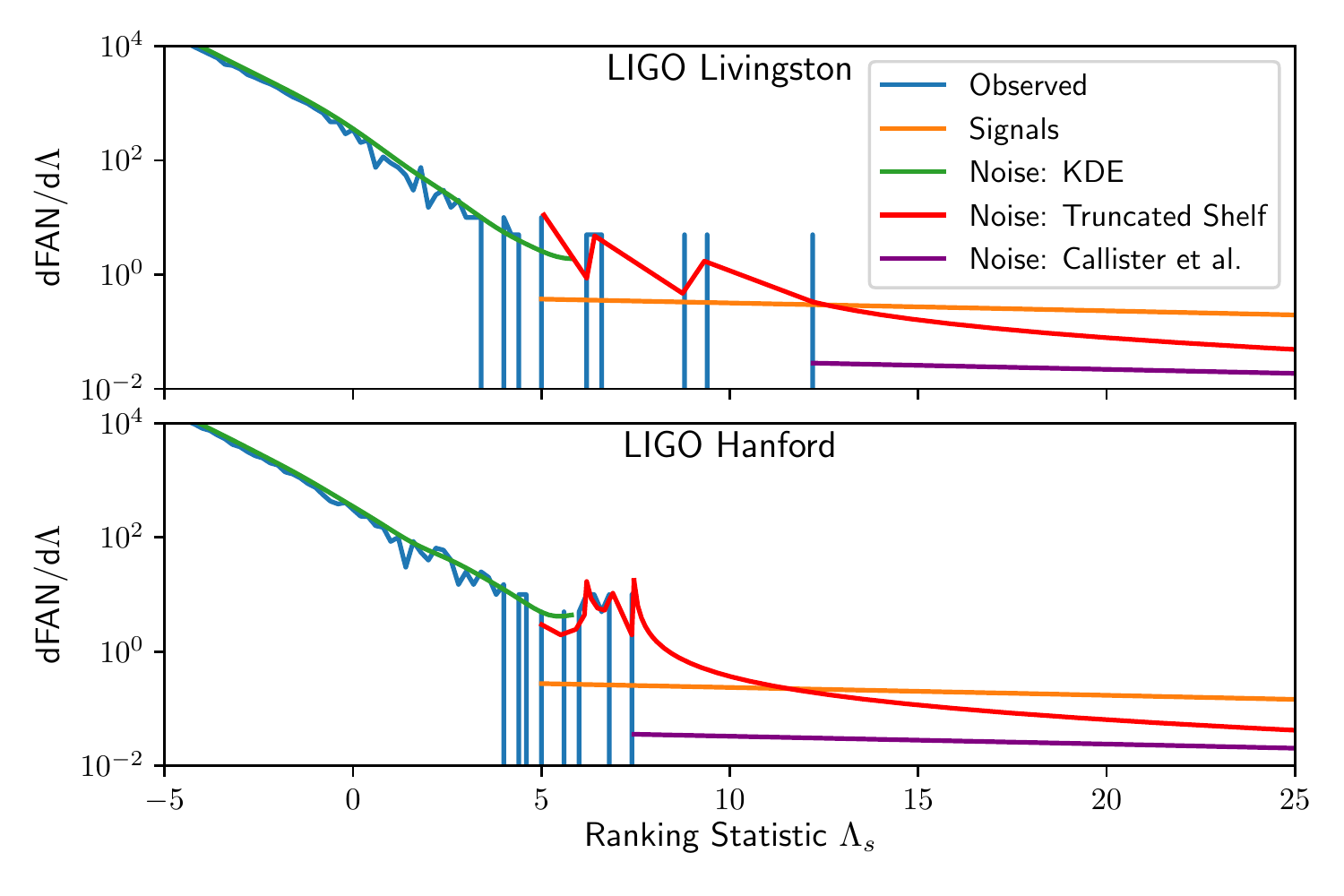}
    \caption{The density of observed single-detector candidates (blue) per observation time as a function of ranking statistic $\Lambda_s$ from the LIGO-Livingston (top) and LIGO-Hanford observatories (bottom). Previously known, confident mergers are removed. Astrophysical significance is determined by comparing the 'truncated shelf' noise model we apply in this paper (red) to the signal model (orange) which was fitted to a simulated population of merger sources. For reference, the \citet{Callister:2017urp} suggested curve is shown in purple.
    }
    \label{fig:signalnoise}
\end{figure}
In practice, real noise is more complicated than our toy example of a simple distribution without astrophysical signals, and where only a single candidate is well separated from the bulk distribution. For our `truncated shelf' model, the form of the noise rate $\mu_N$ is determined by an empirically measured distribution of simulated signals; the normalisation of $\mu_N$ is determined by the $\Lambda_s$ values of both the candidate under consideration and the next candidate in decreasing order.
We apply this model to our highest ranked candidates with $\Lambda_s > 5$.
Fig.~\ref{fig:signalnoise} demonstrates this procedure as applied for the observed single-detector candidates in our BBH search. A natural future extension would be to transition this method to the region where the noise density is high enough that averaging or KDE methods are appropriate.

To reduce the effect of astrophysical contamination in this analysis, we exclude all candidates which were previously identified by a coincident search~\citep{LIGOScientific:2018mvr,Venumadhav:2019tad,Nitz:2019hdf}. As instrument sensitivity and the expected rate of observations increases, the effect of astrophysical contamination also grows. Our analysis of the noise density implicitly considers candidates quieter than a candidate of interest to be predominantly noise; however with a sufficient signal rate, this will no longer be a good approximation. A practical strategy in this case is to only use single-detector candidates from time when multiple comparable instruments are observing to constrain the noise model, under the assumption that these times would have the same noise characteristic as the remaining time. Existing methods for coincident analysis could identify and remove the great majority of detectable astrophysical sources. If this approach had been taken in this work, the result would not have substantially changed, as the distribution of candidates in single-detector time were not significantly different from that in coincident time.

\subsection{Incorporating Additional Detectors}

For candidates that occur when multiple gravitational-wave detectors are observing but have SNR above a threshold of interest in only a single detector, we can update the probability of astrophysical origin we've derived from the data of a single detector (\pastros) using data from the other detectors to derive a combined probability of astrophysical origin (\pastroc). Using PyCBC Inference \citep{Biwer:2018osg}, 
we calculate the Bayes factor $B_{c/s}$ between the astrophysical hypothesis that the data contains a coherent multi-detector gravitational-wave source described by general relativity, versus a noise model where only the single triggering detector observes a signal-like morphology and the remainder observe no signal. This approach is similar to the approach explored in \citet{Veitch:2008ur,Veitch:2009hd}. The use of Bayes factor in searches has also been explored in \citet{Lynch:2015yin,Isi:2018vst}. The Bayes factors are calculated using thermodynamic integration~\citep{Vousden:2015} and crosschecked against nested sampling~\citep{speagle:2019}. This Bayes factor takes into account the possible reduction in phase space (e.g.\ restricting to blind spots of a detector) required for a true signal to not be observable by some detectors. We use the same signal priors as \citet{Nitz:2019hdf} (uniform in co-moving volume, isotropic in sky location and spin orientation, uniform in source-frame component mass) and employ the IMRPhenomPv2 model \citep{Hannam:2013oca,Schmidt:2014iyl}.

We combine $B_{c/s}$ with the single-detector \pastros{} under the assumption that noise is uncorrelated between instruments (a standard choice used by gravitational-wave background estimation) and data in the additional detectors is dominated by the Gaussian noise background with a possible added signal, but not by rare noise artefacts during the time used.
The combined odds $\mathcal{O}_c$ are given by
\begin{equation}
    \mathcal{O}_c = B_{c/s}\mathcal{O}_s, 
\end{equation}
where the astrophysical vs.\ noise odds ratio $\mathcal{O} \equiv p_{\rm astro}/(1 - p_{\rm astro})$. Inverting to obtain \pastroc{} in terms of \pastros{}, we find
\begin{equation}
    \pastroc = \frac{1}{1 + \frac{1 - \pastros}{B_{c/s}\pastros}}.
\end{equation}
For candidates which have large SNR in multiple detectors, a standard time-shifted coincidence background estimate may be preferred as a basis for the probability of astrophysical origin, as it can account for cases where non-Gaussian noise is present. Alternatively, the Bayes factor could be extended to include additional noise models (see for instance \citet{Veitch:2008ur, Veitch:2009hd}).

\section{Binary Black Hole Candidates}

\begin{table*}[t!]
  \begin{center}
    \caption{Candidate events from the search for binary black holes from single-detector observation. Candidates are sorted according to their ranking statistic and separated by detector. Gravitational-wave mergers that were previously identified with high confidence are indicated with checkmarks; for each previously known candidate, \pastros\ is calculated as if that candidate were of unknown origin. For candidates which appear in more than one detector, we calculate $B_{c/s}$ using a
    a noise model where only the detector with the highest SNR contains a (putative) signal. 
    Parameter estimates are in the source frame and reported with $90\%$ credible intervals.}
    \label{table:bbh}
\begin{tabular}{llllllclllr}
Date designation & GPS time & Obs & Known & \rankingstat & \pastros & ln\bayes &\pastroc & \oddsc & \mchirpsrc & $\chieff$\\ \hline
\multicolumn{11}{|c|}{LIGO-Hanford}\\ \hline
170729+07:37:25UTC & 1185349063.74 & HL & -  & 6.68 & 0.05 & $-4.2\pm0.2$ & $<$.01 & $<$.01 & $25.0^{+6.0}_{-5.5}$ & $-0.3^{+0.3}_{-0.4}$ \\ 
151222+04:03:03UTC & 1134792200.18 & HL & -  & 6.80 & 0.03 & $-3.9\pm0.1$ & $<$.01 & $<$.01 & $38.0^{+13.9}_{-9.3}$ & $-0.2^{+0.4}_{-0.4}$ \\ 
170724+03:01:23UTC & 1184900501.59 & HL & -  & 6.90 & 0.02 & $-7.2\pm0.1$ & $<$.01 & $<$.01 & $51.4^{+7.5}_{-15.2}$ & $-0.2^{+0.3}_{-0.4}$ \\ 
151225+04:11:44UTC & 1135051921.02 & H & -  & 7.40 & 0.12 & - & 0.12 & 0.13 & $34.0^{+7.1}_{-5.7}$ & $-0.3^{+0.4}_{-0.3}$ \\ 
170106+11:03:33UTC & 1167735831.93 & HL & -  & 7.46 & 0.01 & $-6.4\pm0.1$ & $<$.01 & $<$.01 & $36.0^{+5.5}_{-8.2}$ & $-0.0^{+0.3}_{-0.4}$ \\ 
170104+10:11:58UTC & 1167559936.60 & HL & \checkmark & 9.21 & 0.30 & $37.0\pm0.2$ & $>$0.99 & $>$100 & $21.0^{+2.0}_{-1.6}$ & $-0.1^{+0.2}_{-0.2}$ \\ 
170814+10:30:43UTC & 1186741861.54 & HLV & \checkmark & 9.34 & 0.32 & $42.0\pm0.3$ & $>$0.99 & $>$100 & $24.3^{+1.4}_{-1.3}$ & $0.1^{+0.1}_{-0.1}$ \\ 
151226+03:38:53UTC & 1135136350.65 & HL & \checkmark & 15.32 & 0.64 & $22.0\pm0.2$ & $>$0.99 & $>$100 & $8.8^{+0.3}_{-0.3}$ & $0.2^{+0.2}_{-0.1}$ \\ 
170608+02:01:16UTC & 1180922494.49 & HL & \checkmark & 21.58 & 0.75 & $28.0\pm0.3$ & $>$0.99 & $>$100 & $8.0^{+0.2}_{-0.2}$ & $0.1^{+0.2}_{-0.1}$ \\ 
150914+09:50:45UTC & 1126259462.43 & HL & \checkmark & 57.72 & 0.86 & $81.6\pm0.3$ & $>$0.99 & $>$100 & $28.1^{+1.7}_{-1.5}$ & $-0.0^{+0.1}_{-0.1}$ \\

\hline
\multicolumn{11}{|c|}{LIGO-Livingston}\\ \hline
170121+21:25:36UTC & 1169069154.58 & HL & \checkmark & 6.52 & 0.04 & $8.7\pm0.2$ & $>$0.99 & $>$100 & $24.9^{+4.3}_{-3.2}$ & $-0.1^{+0.2}_{-0.3}$ \\ 
170402+21:51:50UTC & 1175205128.59 & HL & -  & 6.69 & 0.09 & $-1.2\pm0.1$ & 0.03 & 0.03 & $21.7^{+16.0}_{-5.0}$ & $0.6^{+0.2}_{-0.6}$ \\ 
170608+02:01:16UTC & 1180922494.49 & HL & \checkmark & 6.96 & 0.09 & $28.0\pm0.3$ & $>$0.99 & $>$100 & $8.0^{+0.2}_{-0.2}$ & $0.1^{+0.2}_{-0.1}$ \\ 
170817+03:02:46UTC & 1186974184.74 & HLV & -  & 8.75 & 0.41 & $-0.2\pm0.1$ & 0.36 & 0.57 & $43.1^{+9.5}_{-7.7}$ & $0.2^{+0.2}_{-0.3}$ \\ 
151124+09:30:44UTC & 1132392661.24 & HL & -  & 9.32 & 0.16 & $-6.1\pm0.1$ & $<$.01 & $<$.01 & $39.7^{+10.1}_{-12.6}$ & $0.3^{+0.3}_{-0.5}$ \\ 
160104+12:24:17UTC & 1135945474.38 & L & -  & 12.21 & 0.47 & - & 0.47 & 0.90 & $32.7^{+6.6}_{-8.2}$ & $0.2^{+0.3}_{-0.5}$ \\ 
170823+13:13:58UTC & 1187529256.52 & HL & \checkmark & 12.98 & 0.18 & $16.7\pm0.2$ & $>$0.99 & $>$100 & $28.8^{+4.7}_{-3.2}$ & $0.1^{+0.2}_{-0.2}$ \\ 
170818+02:25:09UTC & 1187058327.08 & HLV & \checkmark & 13.67 & 0.30 & $6.4\pm0.2$ & $>$0.99 & $>$100 & $26.4^{+2.3}_{-2.2}$ & $-0.1^{+0.2}_{-0.3}$ \\ 
170104+10:11:58UTC & 1167559936.60 & HL & \checkmark & 13.75 & 0.31 & $37.0\pm0.2$ & $>$0.99 & $>$100 & $21.0^{+2.0}_{-1.6}$ & $-0.1^{+0.2}_{-0.2}$ \\ 
170809+08:28:21UTC & 1186302519.75 & HLV & \checkmark & 20.12 & 0.68 & $14.3\pm0.2$ & $>$0.99 & $>$100 & $25.0^{+2.4}_{-1.6}$ & $0.1^{+0.2}_{-0.1}$ \\ 
150914+09:50:45UTC & 1126259462.42 & HL & \checkmark & 32.78 & 0.82 & $81.6\pm0.3$ & $>$0.99 & $>$100 & $28.1^{+1.7}_{-1.5}$ & $-0.0^{+0.1}_{-0.1}$ \\ 
170814+10:30:43UTC & 1186741861.53 & HLV & \checkmark & 34.35 & 0.82 & $42.0\pm0.3$ & $>$0.99 & $>$100 & $24.3^{+1.4}_{-1.3}$ & $0.1^{+0.1}_{-0.1}$ \\ 
\end{tabular}

  \end{center}
\end{table*}

The results from our search for BBH candidates are summarized in Table~\ref{table:bbh}. Among the top 15 candidates from each detector we find that 5 (8) are previously identified candidates from standard coincident analyses of LIGO-Hanford (LIGO-Livingston) data. Note that these candidates were
excised from the noise model as described in Sec.~\ref{sec:nm}. For each previously identified candidate, we report the \pastros\ value that would have been assigned if they had been each individually observed as part of the candidate set.
We also note that many of the candidates occurred during multi-detector observing time and are ruled out by their non-observation in more than one detector. There are no new candidates with \pastroc{} above 0.5. 

\begin{figure}[tb!]
    \centering
    \includegraphics[width=\columnwidth]{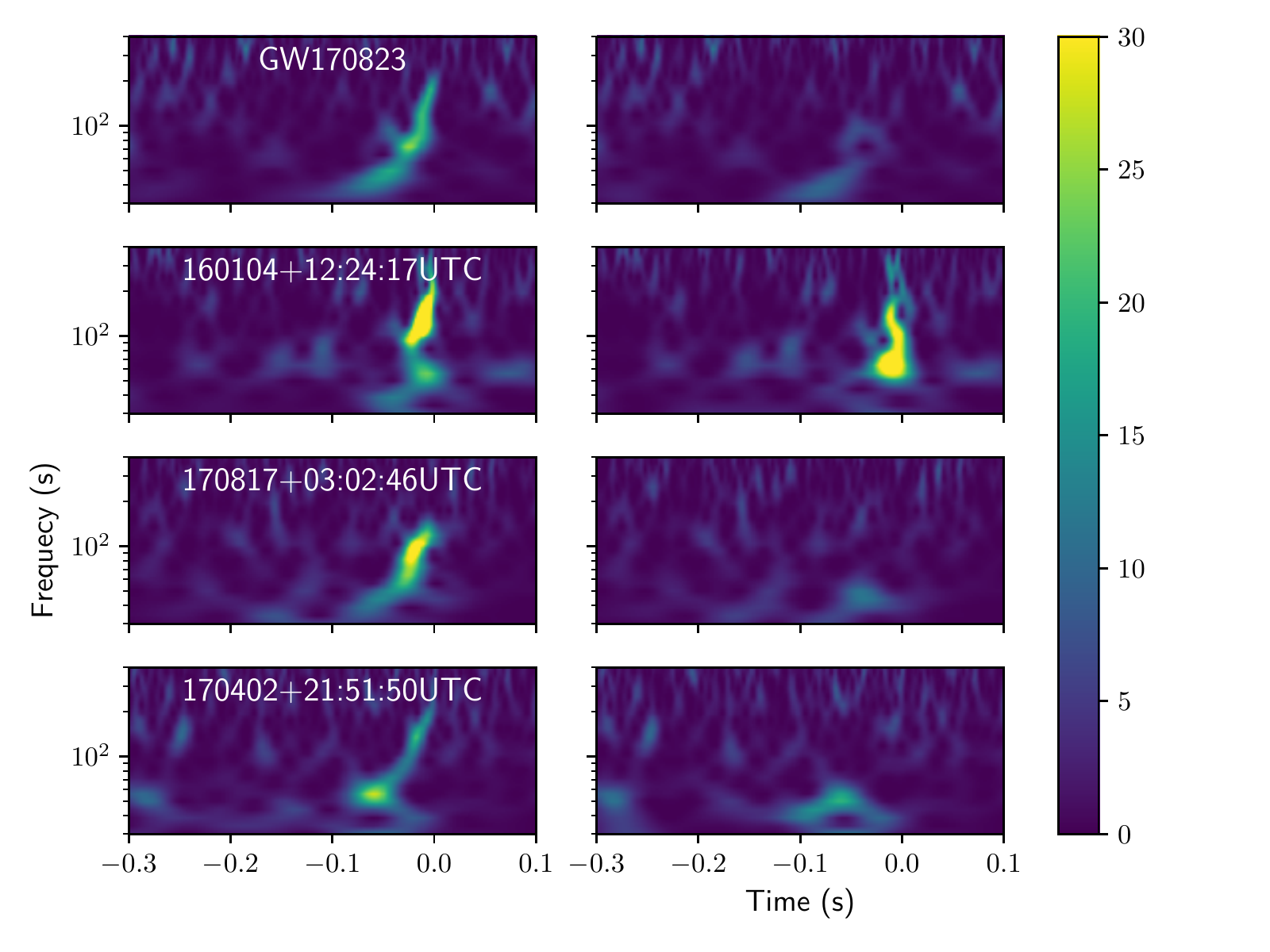}
    \caption{Time-frequency diagnostic plot of the data for select candidates (left) and the same data with best-fit waveform removed (right). The first row is the known observation of GW170823 for comparison. For 160104+12:24:17UTC, we find that best-fit parameters poorly account for the observed data, so we rule out this candidate as a possible gravitational-wave source.}
    \label{fig:qscan}
\end{figure}

For the top candidates, we perform an additional diagnostic to confirm that the signal morphology is consistent with a gravitational-wave source. We take the best-fit parameters for a GR gravitational-wave signal and subtract them from the data. The result for selected candidates is shown in Fig~\ref{fig:qscan}. For some cases, this diagnostic disfavors a candidate due to missing or excess power, such as the case of 160104+12:24:17UTC, which otherwise would have been the most significant candidate. We see that subtracting off the best-fit estimate of the signal introduces visible power at frequencies less than 80 Hz. Our search's standard signal consistency test attempts to exclude such cases, but further improvement in tuning is clearly possible. While human vetting is still important for evaluating candidates~\citep{LIGOScientific:2018mvr,LIGOScientific:2019gag}, a refined test which captures this behavior should be incorporated into future analysis so that it can be naturally accounted for in the determination of $\mu_N$. It may also be possible to compare our candidates with auxiliary channel information, however this is not presently available, and many non-Gaussian noise transients do not have witness channels~\citep{Cabero:2019orq}.

\begin{figure}[tbp!]
    \centering
    \vspace*{-0.6cm}
    \includegraphics[width=\columnwidth]{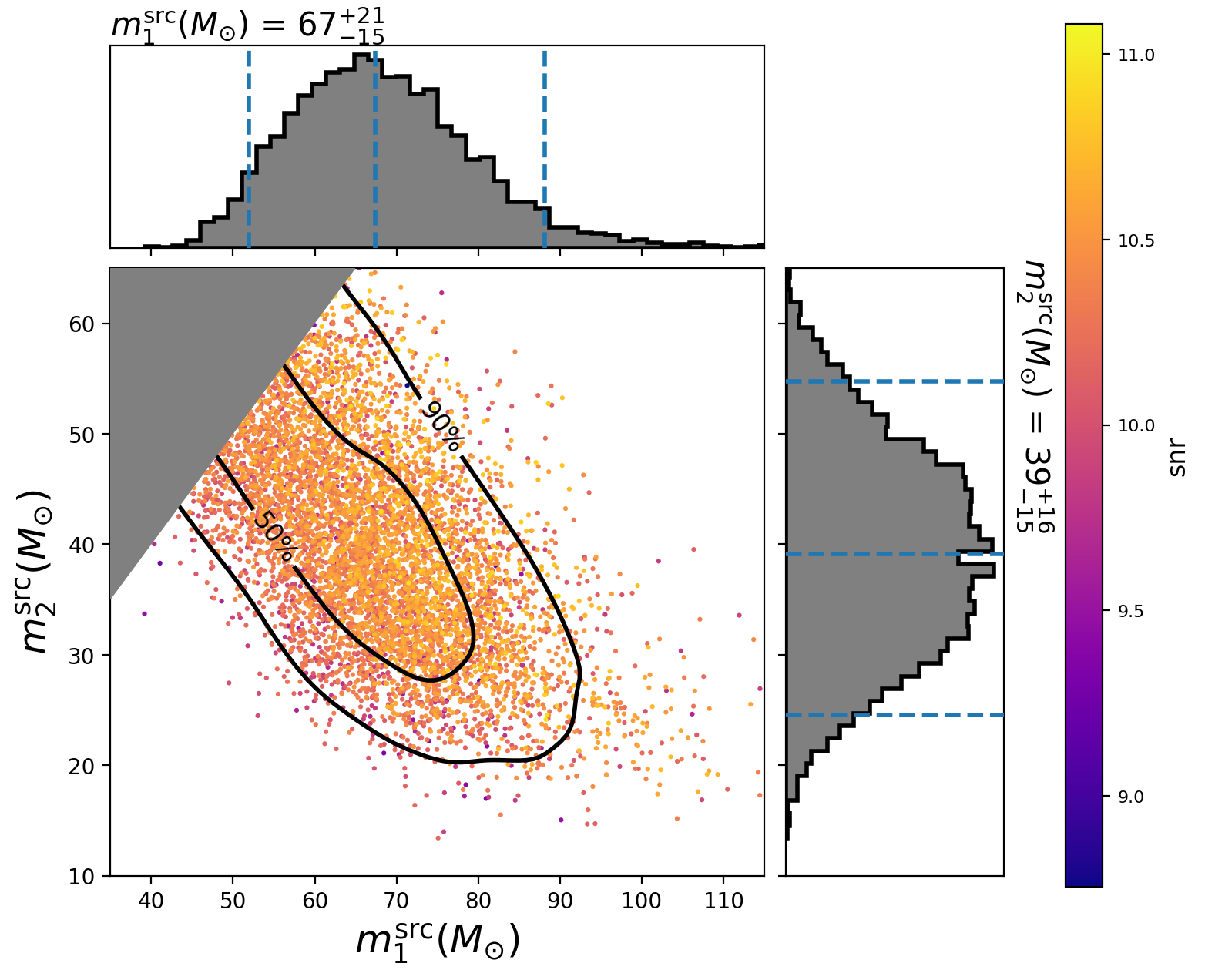}
    \caption{The posterior distribution of the source-frame component masses for 170817+03:02:46UTC with contours of the 50\% and 90\% credible regions. The one-dimensional marginal distributions and their 90\% credible intervals are shown along the axes.}
    \label{fig:pe}
\end{figure}

The most significant remaining candidate from our analysis is 170817+03:02:46UTC with an estimated $\pastroc \sim 0.4$. Alternate choices for the noise and signal models can have a large influence on the estimation of significance. If we directly applied the method of \citet{Callister:2017urp}, we would calculate a $\pastroc \sim 0.8$ if normalising the noise counts to $1$, or 0.6 if normalising to $3$ as originally suggested. One possible KDE extrapolation would give $\pastroc \sim 0.9$, though this is strongly dependent on the chosen bandwidth and kernel. 170402+21:51:50UTC and 170817+03:02:46UT were previously identified as possible mergers in \citet{Zackay:2019btq} with $\pastroc \sim 0.68$ and $\sim 0.86$ respectively, where a noise model more similar
to \citet{Callister:2017urp} was used, as well as different estimations of the signal rate and incorporation of additional detectors. We would obtain comparable results to~\cite{Zackay:2019btq} for these two candidates if we had considered only the same multi-detector observing time and applied their methodology.
Our analysis uses a more conservative noise model;  additionally, for 170402+21:51:50UTC we find that $B_{c/s}$ disfavors the astrophysical hypothesis by 3:1. 

If 170817+03:02:46UTC is astrophysical, it would be consistent with a hierarchical BBH merger, i.e.\ with one or more component being the product of an earlier merger. Fig~\ref{fig:pe} shows the posterior distribution of component masses. 
The primary mass is constrained to $>52\,\msun$ at $95\%$ confidence,
above the cutoff suggested by pair instability for systems which form through standard stellar evolution \citep{Woosley:2016hmi,Belczynski:2016jno,Marchant:2018kun,Woosley_2019,Stevenson:2019rcw}. Support for similar systems in upcoming observing runs would lend credence to this formation channel.

\begin{table*}[tb!]
  \begin{center}
    \caption{Candidate events from the search for binary neutron star mergers from single-detector observation. Candidates are sorted according to their ranking statistic and separated by detector. Parameters are those of the best matching template from our search.}
    \label{table:bns}
\begin{tabular}{lllllll}
Date designation & GPS time & Obs & Known & \rankingstat & $\mathcal{M}^{t}$ & $\chieff^t$\\ \hline
\multicolumn{7}{|c|}{LIGO-Hanford}\\ \hline
170429+21:29:20UTC & 1177536578.86 & HL & - & 4.66 & 1.25 & -0.0 \\ 
170722+20:10:17UTC & 1184789435.61 & H & - & 4.81 & 1.45 & 0.2 \\ 
170630+12:39:04UTC & 1182861562.20 & HL & - & 5.15 & 1.41 & -0.1 \\ 
151222+04:05:43UTC & 1134792360.30 & HL & - & 5.40 & 1.42 & 0.2 \\ 
170817+12:41:04UTC & 1187008882.45 & HLV & \checkmark & 55.28 & 1.20 & -0.0 \\ 

\hline
\multicolumn{7}{|c|}{LIGO-Livingston}\\ \hline
170320+19:44:35UTC & 1174074293.10 & HL & - & 4.87 & 1.25 & -0.0 \\ 
151013+10:40:09UTC & 1128768026.92 & L & - & 4.91 & 1.46 & -0.0 \\ 
170723+20:49:13UTC & 1184878171.60 & HL & - & 4.99 & 1.17 & -0.0 \\ 
151031+01:46:02UTC & 1130291179.17 & L & - & 5.24 & 1.40 & 0.1 \\
170817+12:41:04UTC & 1187008882.44 & HLV & \checkmark & 82.23 & 1.20 & -0.0 \\ 
\end{tabular}
  \end{center}
\end{table*} 

\section{Binary Neutron Star Candidates}

We present the most significant BNS candidates in Table~\ref{table:bns}. A similar procedure as done for BBH candidates could be applied to calculate astrophysical significance, however, none of these candidates significantly depart from the expected noise background, except for the clear detection of GW170817. The next-loudest candidate after GW170817, 151222+04:05:43UTC, has $\pastros \sim .008$. Due to the low expected signal rate of this data set, a very loud signal could at most approach a \pastros\ of 0.5. Although we have not calculated \pastroc\ or $B_{c/s}$ for these candidates, as the sensitivity of the gravitational-wave network improves, and the expected rate of BNS mergers increases, the methodology we've applied to BBH detections will similarly apply here. Since the calculation of evidences for long-duration signals is more computationally intensive than short-duration BBH signals, the development of fast or approximate methods to calculate $B_{c/s}$ would aid in applying our methods to a larger candidate set.

\section{Significance of GW190425}

From information in \citet{Abbott:2020uma}, we can see that GW190425 was the most significant BNS candidate observed in O1-O3 up to April 2019, excluding previously known candidates. Plots therein indicate the candidate's ranking statistic is well separated from the existing background; also, PyCBC-based search results are given using a re-weighted SNR ranking statistic \citep{Usman:2015kfa,Nitz:2017svb}. This allows us to estimate the signal distribution by assuming the ranking statistic $\hat{\rho}$ would follow $\hat{\rho}^{-4}$ distribution \citep{Schutz:2011tw}. 
If we apply the method developed here, we find that the signal is sufficiently loud that the method of \citet{Callister:2017urp} and ours give similar results,
in which case the astrophysical significance reduces to 
\begin{equation}
    p_{190425,s} = \frac{R(S>\rho*)} {R(S>\rho*) + R(N>\rho*)},
\end{equation}
where $R(N>\rho*)$ is the rate of noise candidates per the total observing time with comparable or greater ranking statistic (taken to be ${\sim}1$), and $R(S>\rho*)$ is the expected rate of detections ($\sim 2$), which we take from the observed time, average instrument sensitivity, and previously estimated rate, yielding $p_{190425,s}\sim 0.7$. Taking into account the data in Virgo, we find that $\ln B_{c/s} \sim -0.7\pm 0.3$, yielding $p_{190425, c} \sim 0.5$. This probability is sensitive to both the noise and signal rate estimates. With only a handful of observations, we can expect the detected merger rate to have order of magnitude uncertainties. Furthermore, this estimate has implicitly assumed that GW170817 and GW190425 are of the same class, which is unclear given the unusually high mass observed for GW190425 compared to galactic BNS systems \citep{Abbott:2020uma}. Further observations will reduce the uncertainty in expected rate. Note that we cannot directly compare our probability to the false alarm rate of 1 per 69,000 years stated in \citet{Abbott:2020uma}: formally, only a 1 per observation time false alarm rate can be measured empirically and the lower rate estimate results from KDE extrapolation \citep{Messick:2016aqy}.

We note that additional followup analyses also contributed to validation of GW190425 as an event consistent with a GW signal \citep{Abbott:2020uma}, though their effect on statistical measures of its significance may be hard to quantify. 

\section{Conclusions}

We have presented a single-detector search method and analyzed the 115 days of single-detector LIGO public data for gravitational-waves from BBH and BNS mergers.  While we find no candidates during this time with $\past > 0.5$, the high observed rate of binary mergers indicates that this kind of detection will not be uncommon in the future, assuming a non-negligible fraction of single-detector observing time, or time when only one detector is operating at high sensitivity. Our full analysis also searches the complete set of LIGO data for single-detector candidates. For candidates which occur when multiple detectors are observing, we introduce a method to update the single-detector probability of astrophysical origin based on support from additional detectors' data.

170817+03:02:46UTC ($\pastroc \sim 0.4$) is the top candidate from our full analysis. The candidate occurred when Virgo and the two LIGO detectors were operating and is consistent with a hierarchical merger if astrophysical. The probability of astrophysical origin for a given candidate is highly dependent on the noise model. We choose an approach similar to the one proposed in \citet{Callister:2017urp} modified to provide more robust statements in the case of marginal candidates.

Our search identifies 10 out of the 14 BBH mergers in the 2-OGC catalog \citep{Nitz:2019hdf} within the top 25 candidates in either LIGO-Hanford or LIGO-Livingston. This suggests it may be possible to build an effective search by following up single-detector candidates. Existing compact-binary search methods as employed by PyCBC \citep{Usman:2015kfa,Davies:2020} and others \citep{Messick:2016aqy,Adams:2015ulm} are not currently constrained by their ability to estimate background, however, as analysis methods become more sophisticated, it may be useful to investigate this followup approach further.

Posterior samples and candidate lists are available in the associated data release (\release).

\acknowledgments
We thank Tom Callister, Ryan Magee, Tejaswi Venumadhav, and Barak Zackay for their comments and suggestions. We acknowledge the Max Planck Gesellschaft and the Atlas cluster computing team at AEI Hannover for support. TD and GSD acknowledge support from the Maria de Maeztu Unit of Excellence MDM-2016-0692 and Xunta de Galicia. This research has made use of data, software and/or web tools obtained from the Gravitational Wave Open Science Center (https://www.gw-openscience.org), a service of LIGO Laboratory, the LIGO Scientific Collaboration and the Virgo Collaboration. LIGO is funded by the U.S. National Science Foundation. Virgo is funded by the French Centre National de Recherche Scientifique (CNRS), the Italian Istituto Nazionale della Fisica Nucleare (INFN) and the Dutch Nikhef, with contributions by Polish and Hungarian institutes.
\bibliography{references}

\end{document}